\documentclass[a4,11pt]{cip-submit}  
\usepackage{afterpage}
\usepackage{mathtools,slashed}

\usepackage{mathptmx}
\usepackage{hyperref}
\hypersetup{
colorlinks=false,
pdfborder={0 0 0}
}
\def\Mr{\uppercase}
\usepackage{float}
\usepackage{wasysym}
\usepackage{graphicx,wrapfig,tikz}
\usepackage{caption}
\usepackage{subcaption}
\usepackage{amsmath,amsxtra,
amssymb,latexsym,amscd,color,bm}
\usepackage[mathscr]{eucal}
\usepackage{epstopdf}
\usepackage{array}
\usepackage{cite}

\def\vsm{\vskip0.1cm}
\def\doi#1{\href{http://dx.doi.org/#1}{doi:#1}}

\def\titles#1{\title{\large\bf\noindent #1}}
\def\authors#1{\author{\begin{flushleft}{#1}\end{flushleft}}}
\def\authord#1#2{\indent\Mr{#1}\\
	\textit{\indent#2}\vsm}

\def\email#1{\bigskip\href{mailto:#1}{\textit{E-mail:}~{#1}}\\[3mm]}

\def\received#1{\vsm\textit{\indent Received #1}}
\def\accepted#1{\vsm\textit{Accepted for publication~#1}}

\def\and{$\text{\tiny AND }$}

\begin{document}
\titles{One-loop off-shell decay 
$H^* \rightarrow ZZ$ at future colliders}
\authors{\authord{Khiem Hong Phan, Dzung Tri Tran}
{
Institute of Fundamental and Applied Sciences, 
Duy Tan University, Ho Chi Minh City $700000$, Vietnam\\ 
Faculty of Natural Sciences, Duy Tan University, 
Da Nang City $550000$, Vietnam
}
\authord{Anh Thu Nguyen}
{
University of Science Ho Chi Minh City, $227$ 
Nguyen Van Cu, District $5$, Ho Chi Minh City, Vietnam
}
\email{phanhongkhiem@duytan.edu.vn}
\received{\today}
\accepted{DD MM YYYY}  
}
\maketitle
\markboth{Khiem Hong Phan }
{One-loop off-shell decay 
$H^* \rightarrow ZZ$ at future colliders}
\begin{abstract} 
We present one-loop formulas for
contributing to the 
$HZZ$ vertex in 't Hooft-Veltman 
gauge within Standard Model 
framework. One-loop off-shell Higgs 
decay rates to $Z$-pair 
are investigated in both 
unpolarized  
and longitudinal polarization 
for $Z$ bosons in final state.
The corrections are range of $7\%$ 
to $8.4\%$ when we vary the 
off-shell Higgs mass from $200$ GeV to 
$500$ GeV. In applications, we study 
off-shell Higgs decay $H^* \rightarrow ZZ$
in the Higgs productions at 
future colliders such as the signal 
processes $\gamma^*(Q^2)\gamma 
\rightarrow H^* \rightarrow ZZ$
and 
$e^-\gamma 
\rightarrow e^-H^* \rightarrow e^- ZZ$
are analyzed. 
\end{abstract}
\section{Introduction}
Since the discovery of the 
Standard-Model-like (SM-like) Higgs boson 
at the Large Hadron Collider 
(LHC)~\cite{ATLAS:2012yve,CMS:2012qbp},
High energy physics have entered a new era. 
High-precision measurements of the Higgs 
properties are the top priority tasks at 
the LHC. So far, most of the measurements 
at the LHC 
focus on the on-shell Higgs productions and 
on-shell Higgs decay channels. The data 
shows that the Higgs signal strengths 
are in agreement with the SM predictions. 
The above measurements are planned
to probe as precisely as possible in 
near future
for the high-precision tests of 
the SM as well as extracting 
new physics beyond the SM (BSM). 
Besides that, in order to explore the nature 
of Higgs sector at different energy scales, 
off-shell Higgs 
decay channels are considerable interests. 
Recently, off-shell Higgs decay
$H^*\rightarrow Z^*Z^* \rightarrow 4$ 
leptons have been 
measured at the LHC in Refs.\cite{CMS:2014quz,
ATLAS:2015cuo,ATLAS:2015pre,ATLAS:2018jym,CMS:2019ekd,
CMS:2021ziv}.

We argue that off-shell Higgs decay channel 
$H^*\rightarrow ZZ \rightarrow 4$ leptons
provides rich of phenomenological investigations.
In Ref.\cite{Lee:2018fxj}, examining the tail of 
off-shell Higgs decay mode $H^*\rightarrow Z_LZ_L$, 
one can test the unitarity of the SM and explore 
new physics through high energy behavior. 
Searching for new physics 
through off-shell Higgs decay
$H^*\rightarrow ZZ \rightarrow l^+l^-\nu_l\bar{\nu}_l$ 
has been studied in Ref.\cite{Goncalves:2020vyn}
in which the authors have considered
the effective theory by proposing the energy-dependence
operators. Off-Shell Higgs decays as a probe of naturalness
have been discussed in Ref.\cite{Goncalves:2017iub}
and as a probe of the trilinear Higgs coupling
have been reported in Ref.\cite{Haisch:2021hvy}. 
Through the examination of the decay channel 
$H \rightarrow ZZ^* \rightarrow 4$
leptons at the LHC,
the authors in Ref.~\cite{Godbole:2007cn}
have been studied
the effects of the CP-conserving 
and CP-violating of the
general $HZZ$ coupling. 
Other phenomenological studies of 
off-shell Higgs decay in many of BSMs
have been found in Refs.\cite{Azatov:2022kbs,
Cacciapaglia:2014rla,Logan:2014ppa,Englert:2014ffa,
Chen:2015iha,
Goncalves:2018pkt,Dwivedi:2016xwm}. 

At the LHC, one-loop and two-loop
QCD corrections
for both signal and background of 
the off-shell Higgs decay
$H^*\rightarrow ZZ$  
have been computed in~\cite{Caola:2015ila, 
Caola:2015psa,Campbell:2016ivq,Caola:2016trd,Grober:2019kuf,
Davies:2020lpf,Alioli:2021wpn,Grazzini:2021iae,Buonocore:2021fnj,
Haisch:2022rkm}. One-loop electroweak corrections to 
Higgs boson decay into $ZZ$  in standard model 
have been reported in 
~\cite{Kniehl:1990mq}, in 
the minimal supersymmetric 
model~\cite{Pierce:1992hg, Hollik:2011xd}, 
and to Higgs boson decay into four leptons 
have been evaluated in 
Refs.~\cite{Boselli:2015aha,Bredenstein:2006rh,
Bredenstein:2006ha}. 
Off-shell Higgs decay effects in Higgs 
productions at future linear 
colliders have been studied in 
Refs.~\cite{Liebler:2015aka,Yan:2021tmw} 
in which
the authors have been included only
the tree level vertex $HZZ$ 
in the analysis. Due to the important roles 
of the off-shell Higgs decays 
and in order 
to match the high-precision data at 
future colliders, we evaluate
for one-loop electroweak corrections to 
the off-shell decay $H^* \rightarrow ZZ$
in this work. The computation is performed in 
't Hooft-Veltman gauge. Analytic 
formulas for one-loop form factors 
in the decay process are expressed 
in terms of Passarino-Veltman scalar functions  
in the standard notations of {\tt LoopTools}.  
As a result, the off-shell decay rates
can be evaluated numerically by using 
this package.
One-loop electroweak corrections 
to the off-shell decay
rates are studied for the cases of
unpolarized $Z$ bosons and longitudinal
polarization of $Z$ bosons in final state.
In applications, we study 
off-shell Higgs decay $H^* \rightarrow ZZ$
in the Higgs productions at 
future colliders such as the signal 
processes $\gamma^*(Q^2)\gamma 
\rightarrow H^* \rightarrow ZZ$
and 
$e^-\gamma 
\rightarrow e^-H^* \rightarrow e^- ZZ$
are analyzed.

The layout of the paper is as follows: 
In section $2$, we present 
one-loop expressions for the 
vertex $HZZ$. Phenomenological
results for this work are  
shown in the section $3$. 
Conclusions
and outlook for this research 
are discussed in the 
section $4$. 
\section{Calculations} 
In general, one-loop contributions to 
the $H(p)Z(q_1)Z(q_2)$ 
vertex are decomposed in terms of 
Lorentz structure as follows:
\begin{eqnarray}
 \mathcal{V}_{HZZ}^{1-\text{loop}}
 (p, q_1, q_2)
 &=& g_{HZZ} \Big( F_{00}\; g^{\mu\nu} 
 +\sum\limits_{i,j = 1}^2 F_{ij}\; 
 q_i^{\mu}q_j^{\nu}
 \Big).
\end{eqnarray}
The tree level coupling of the
Higgs to $ZZ$ is given
$g_{HZZ} = \frac{e M_W}{c_W^2 s_W}$
where $s_W$ and $c_W$ are sine and cosine 
of Weinberg angle respectively. In this 
expression, the terms 
$F_{00}$, $F_{ij}$ for $i,j=1,2$ are 
denoted for one-loop form factors. 
These form factors are functions 
of $p^2, q_1^2, q_2^2$ and they are expressed
in terms of Passarino-Veltman scalar functions 
(called as PV-functions hereafter). All one-loop 
Feynman diagrams contributing to 
this vertex can be grouped into three classes 
as follows (shown in appendix $D$). By 
considering all fermions exchanging in the 
loop diagrams, this is corresponding to
group $1$. With including all $W$ bosons, 
goldstone bosons and ghost particles 
propagating in the loop, 
we have group $2$ accordingly. One finally 
takes $Z$ boson, Higgs and goldstone 
bosons exchanging in the loop, we have 
correspondingly to group $3$. It is known 
that one-loop
contributing to the vertex $HZZ$ contains 
ultraviolet divergent ($UV$-divergent).
Following renormalization theory, 
the counter-terms are given for 
cancelling the $UV$-divergent. 
We then have counter-term diagrams
in group $0$ which their analytic 
formulas are presented in appendix $C$.

Analytic results for the above form factors
are computed class by class of Feynman diagrams
as follows. First, one-loop amplitudes 
for all Feynman diagrams mentioned in above 
are written down. 
We then handle with Dirac traces and Lorentz 
contractions in $d$ dimensions by using 
{\tt Package-X}~\cite{Patel:2015tea}. The amplitudes 
are next decomposed into tensor one-loop integrals. 
By following tensor reduction for 
one-loop integrals in~\cite{Denner:2005nn}, 
the tensor integrals are then expressed in terms of
the PV-functions which they can be evaluated numerically
by using {\tt LoopTools}~\cite{Hahn:1998yk}. In detail,
analytical results for all form factors
are shown in the following paragraphs. General expression
for the form factor
$F_{00}(p^2; q_1^2, q_2^2)$ is written
as follows:
\begin{eqnarray}
F_{00}(p^2; q_1^2, q_2^2) &=&\sum\limits_{G=\{G_{0}, 
G_1, \cdots, G_3\}}F_{00}^{(G)}(p^2; q_1^2, q_2^2). 
\end{eqnarray}
Where $\{G_{0}, G_1, \cdots, G_3\} = 
\{\text{group 0}, 
\text{group 1} \cdots, \text{group 3}\}$ are groups 
of Feynman diagrams showing in appendix $D$.
The form factor $F_{00}^{(G_0)} (p^2; q_1^2, q_2^2)$
for group $0$
is expressed in the appendix $C$. 
In group $1$, we consider
all fermions exchanging in the loop. 
We take top quark in the loop for 
an example. The resulting for 
the form factor reads:
\begin{eqnarray}
F_{00}^{(G_1)} (p^2; q_1^2, q_2^2)
&=&  
\dfrac{e^3}{576\pi^2 M_W s_W^3 c_W^2} 
N^C_t m_t^2
\times
\\
&&
\hspace{-0.7cm}
\times
\Bigg\{
2 (32 s_W^4-24 s_W^2+9) 
B_0(p^2,m_t^2,m_t^2)
+9 \Big[
B_0(q_1^2,m_t^2,m_t^2)
+ B_0(q_2^2,m_t^2,m_t^2)
\Big]
\nonumber\\
&&\hspace{-0.7cm}
+ \Big[36 m_t^2
+ 8 s_W^2 (3 - 4 s_W^2) 
(-p^2+q_1^2+q_2^2)
-9 (q_1^2+q_2^2)
\Big]
C_0(p^2,q_1^2,q_2^2,m_t^2,m_t^2,m_t^2)
\nonumber\\
&&
\hspace{-0.7cm}
-8(32 s_W^4-24 s_W^2+9) 
C_{00}(p^2,q_1^2,q_2^2,m_t^2,m_t^2,m_t^2)
\Bigg\}.\nonumber
\end{eqnarray}
Here $N^C_t=3$ is color number top quark.
For group 2, we take into account 
all $W$ boson in the loop diagrams. 
The form factor is then given:
\begin{eqnarray}
F_{00}^{(G_2)}(p^2; q_1^2, q_2^2) 
&=& 
\dfrac{e^3}{64 \pi^2 M_W s_W^3 c_W^2}
\times
\\
&&\times
\Bigg\{
4 M_W^2 (c_W^4-s_W^4)
\Big[ B_0(q_1^2,M_W^2,M_W^2)
+ B_0(q_2^2,M_W^2,M_W^2) \Big]
\nonumber\\
&&
- 4 M_W^2 \Big[c_W^4 
\Big(5 p^2-4 (q_1^2+q_2^2+M_W^2)\Big)
+ s_W^2 c_W^2 \Big(q_1^2+q_2^2-2 (p^2+M_W^2)\Big)
\nonumber\\
&&
+ s_W^4 \Big(2 M_W^2 + M_H^2\Big)
\Big]
C_0(p^2,q_1^2,q_2^2,M_W^2,M_W^2,M_W^2) 
\nonumber\\
&&
+\Big[4 M_H^2 (c_W^2-s_W^2)^2
+8 M_W^2 \Big(c_W^4 (4 d-7)
-2 c_W^2 s_W^2+s_W^4\Big)
\Big] 
\times
\nonumber\\
&& 
\hspace{6cm}
\times
C_{00}(p^2,q_1^2,q_2^2,M_W^2,M_W^2,M_W^2)
\nonumber\\
&&
- \Big[8 M_W^2 c_W^2 
\big(c_W^2 (d-2)-s_W^2\big)
+ M_H^2 (c_W^2-s_W^2)^2\Big]
B_0(p^2,M_W^2,M_W^2)
\Bigg\}. \nonumber
\end{eqnarray}
We next consider the contributions 
from group $3$ in which the particles 
$Z,\chi_3, H$ are exchanged
in the loop diagrams. 
The form factor is evaluated 
accordingly: 
\begin{eqnarray}
F_{00}^{(G_3)}(p^2; q_1^2, q_2^2) 
&=& 
-\dfrac{e^3}{128 \pi^2 M_W s_W^3 c_W^6}
\Bigg\{
4 c_W^2 M_W^2 
\Big[
B_0(q_1^2,M_H^2,M_Z^2)
+B_0(q_2^2,M_H^2,M_Z^2)
\\
&& 
\hspace{5cm}
+3 M_H^2 C_0(p^2,q_1^2,q_2^2,M_H^2,M_H^2,M_Z^2)
\Big]
\nonumber\\
&&
+ M_H^2 c_W^4 B_0(p^2,M_Z^2,M_Z^2)
+3 c_W^4 M_H^2 B_0(p^2,M_H^2,M_H^2) 
\nonumber\\
&&
+8 M_W^4 C_0(p^2,q_1^2,q_2^2,M_Z^2,M_Z^2,M_H^2)
-12 M_H^2 c_W^4 
C_{00}(p^2,q_1^2,q_2^2,M_H^2,M_H^2,M_Z^2)
\nonumber\\
&&
-4 c_W^2 (c_W^2 M_H^2+2 M_W^2) 
C_{00}(p^2,q_1^2,q_2^2,M_Z^2,M_Z^2,M_H^2)
\Bigg\}
. \nonumber
\end{eqnarray}
Other form factors ($F_{ij}$ for $i,j=1,2$)
are also written in the form of
\begin{eqnarray}
F_{ij}(p^2; q_1^2,q_2^2) &=&\sum\limits_{G=\{ 
G_1, \cdots, G_3\}}F_{ij}^{(G)}(p^2; q_1^2,q_2^2).
\end{eqnarray}
All form factors $F_{ij}$ for $i,j =1,2$
are the UV-finite. Therefore, we have only
group $1$ to group $3$ contributing to 
these form factors.
Applying the same procedure, each 
form factor is expressed as follows: 
\begin{eqnarray}
F_{11}^{(G_1)} (p^2; q_1^2, q_2^2) 
&=& 
-\dfrac{e^3
(32 s_W^4-24 s_W^2+9)
}{144 \pi^2 M_W s_W^3 c_W^2} N^C_t m_t^2 
\times
\\
&&
\times
\Big[
C_1(p^2,q_1^2,q_2^2,m_t^2,m_t^2,m_t^2)
+2 C_{11}(p^2,q_1^2,q_2^2,m_t^2,m_t^2,m_t^2)
\Big]
, 
\nonumber 
\end{eqnarray}
\begin{eqnarray}
F_{11}^{(G_2)}(p^2; q_1^2, q_2^2) 
&=& 
\dfrac{e^3}{32 \pi^2 M_W s_W^3 c_W^2}
\times
\\
&&\times
\Bigg\{
\Big[
M_H^2 (c_W^2-s_W^2)^2 
+M_W^2 \Big(2 c_W^4 (4 d-7)
-3 c_W^2 s_W^2+3 s_W^4\Big)
\Big] \times
\nonumber\\
&& \hspace{5cm}
\times
C_1(p^2,q_1^2,q_2^2,M_W^2,M_W^2,M_W^2)
\nonumber\\
&&
+\Big[
2 M_H^2 (c_W^2-s_W^2)^2
+4 M_W^2 \Big(c_W^4 (4 d-7)
-2 c_W^2 s_W^2+s_W^4\Big)
\Big] 
\times
\nonumber\\
&& 
\hspace{5cm}
\times
C_{11}(p^2,q_1^2,q_2^2,M_W^2,M_W^2,M_W^2)
\nonumber\\
&&
+
2 M_W^2 s_W^2 (s_W^2-2 c_W^2) 
C_0(p^2,q_1^2,q_2^2,M_W^2,M_W^2,M_W^2)
\Bigg\}
, \nonumber 
\end{eqnarray}
\begin{eqnarray}
F_{11}^{(G_3)}(p^2; q_1^2, q_2^2) 
&=& 
\dfrac{e^3}{64 \pi^2 M_W s_W^3 c_W^4}
\Bigg\{
2 M_W^2 C_0(p^2,q_1^2,q_2^2,M_Z^2,M_Z^2,M_H^2)
\\
&& 
+ (c_W^2 M_H^2+3 M_W^2)
C_1(p^2,q_1^2,q_2^2,M_Z^2,M_Z^2,M_H^2)
\nonumber\\
&&
+3 c_W^2 M_H^2
\Big[
C_1(p^2,q_1^2,q_2^2,M_H^2,M_H^2,M_Z^2)
+ 2 C_{11}(p^2,q_1^2,q_2^2,M_H^2,M_H^2,M_Z^2)
\Big]
\nonumber\\
&&
+2 (c_W^2 M_H^2+2 M_W^2) 
C_{11}(p^2,q_1^2,q_2^2,M_Z^2,M_Z^2,M_H^2)
\Bigg\}
. \nonumber
\end{eqnarray}
Analytic results for 
the form factors $F_{12}$ are shown as:
\begin{eqnarray}
F_{12}^{(G_1)} (p^2; q_1^2, q_2^2) 
&=& 
-\dfrac{e^3}{288 
\pi^2 M_W s_W^3 c_W^2} N^C_t m_t^2
\Bigg\{
8 s_W^2 (3-4 s_W^2) 
C_0(p^2,q_1^2,q_2^2,m_t^2,m_t^2,m_t^2)
\nonumber\\
&&
-(3-8 s_W^2)^2 
\Big[
C_1(p^2,q_1^2,q_2^2,m_t^2,m_t^2,m_t^2)
+C_1(p^2,q_2^2,q_1^2,m_t^2,m_t^2,m_t^2)
\Big]
\nonumber\\
&&
-4 (32 s_W^4-24 s_W^2+9) 
C_{12}(q_1^2,p^2,q_2^2,m_t^2,m_t^2,m_t^2)
\Bigg\}
, 
\end{eqnarray}
\begin{eqnarray}
F_{12}^{(G_2)}(p^2; q_1^2, q_2^2) 
&=& 
\dfrac{e^3}{64 \pi^2 M_W s_W^3 c_W^2}
\Bigg\{
\Big[4 M_W^2 
\Big(
3 s_W^2 c_W^2
+s_W^4
-2 c_W^4 (d-2)
\Big)
- M_H^2 (c_W^2-s_W^2)^2\Big]
\nonumber\\
&&
\hspace{2cm}
 \times
C_0(p^2,q_1^2,q_2^2,M_W^2,M_W^2,M_W^2)
\nonumber\\
&&
+\Big[
M_H^2 (c_W^2-s_W^2)^2 
+M_W^2 \Big(2 c_W^4 (4 d-9)
-11 c_W^2 s_W^2
-s_W^4\Big)
\Big]
\times
\nonumber\\
&&
\hspace{0.5cm}
\times 
\Big[
C_1(p^2,q_2^2,q_1^2,M_W^2,M_W^2,M_W^2)
-2 C_1(p^2,q_1^2,q_2^2,M_W^2,M_W^2,M_W^2)
\Big]
\nonumber\\
&&
- 4 \Big[
M_H^2 (c_W^2-s_W^2)^2 
+2 M_W^2 \Big(c_W^4 (4 d-7)
-2 s_W^2 c_W^2
+s_W^4\Big)
\Big] 
\\
&&
\hspace{5.5cm}
\times
C_{12}(q_1^2,p^2,q_2^2,M_W^2,M_W^2,M_W^2)
\Bigg\}
, \nonumber 
\end{eqnarray}
\begin{eqnarray}
F_{12}^{(G_3)}(p^2; q_1^2, q_2^2) 
&=& 
\dfrac{e^3}{128 \pi^2 M_W s_W^3 c_W^4}
\Bigg\{
(4 M_W^2-c_W^2 M_H^2) 
C_0(p^2,q_1^2,q_2^2,M_Z^2,M_Z^2,M_H^2)
\\
&&
-3 c_W^2 M_H^2 
C_0(p^2,q_1^2,q_2^2,M_H^2,M_H^2,M_Z^2)
\nonumber\\
&&
-6 c_W^2 M_H^2 
\Big[
C_1(p^2,q_1^2,q_2^2,M_H^2,M_H^2,M_Z^2)
+C_1(p^2,q_2^2,q_1^2,M_H^2,M_H^2,M_Z^2)
\nonumber\\
&&\hspace{6cm}
+2 C_{12}(q_1^2,p^2,q_2^2,M_Z^2,M_H^2,M_H^2)
\Big]
\nonumber\\
&&
+\Big(2 M_W^2 -2 c_W^2 M_H^2\Big) 
\Big[
C_1(p^2,q_1^2,q_2^2,M_Z^2,M_Z^2,M_H^2)
+C_1(p^2,q_2^2,q_1^2,M_Z^2,M_Z^2,M_H^2)
\Big]
\nonumber\\
&&
- 4 \Big(2 M_W^2 + c_W^2 M_H^2 \Big)
C_{12}(q_1^2,p^2,q_2^2,M_H^2,M_Z^2,M_Z^2)
\Bigg\}
. \nonumber
\end{eqnarray}
We list all analytic expressions for
the form factor  $F_{21}$ as follows:
\begin{eqnarray}
F_{21}^{(G_1)} (p^2; q_1^2, q_2^2) 
&=& 
-\dfrac{e^3}{288 \pi^2 M_W s_W^3 c_W^2} 
N^C_t m_t^2
\times
\\
&&
\times
\Bigg\{
8 s_W^2 (4 s_W^2-3) 
C_0(p^2,q_1^2,q_2^2,m_t^2,m_t^2,m_t^2)
-9 \Big[
C_1(p^2,q_1^2,q_2^2,m_t^2,m_t^2,m_t^2)
\nonumber\\
&&
+C_1(p^2,q_2^2,q_1^2,m_t^2,m_t^2,m_t^2)
\Big]
-4 (32 s_W^4-24 s_W^2+9) 
C_{12}(q_1^2,p^2,q_2^2,m_t^2,m_t^2,m_t^2)
\Bigg\}
, \nonumber
\end{eqnarray}
\begin{eqnarray}
F_{21}^{(G_2)}(p^2; q_1^2, q_2^2) 
&=& 
-\dfrac{e^3}{16 \pi^2 M_W s_W^3 c_W^2}
\Bigg\{ 
\Big[
M_H^2 (c_W^2-s_W^2)^2
+2 M_W^2 
\Big(c_W^4 (4 d-7) + s_W^2 (s_W^2 -2 c_W^2)\Big)
\Big] 
\nonumber\\
&&
\hspace{7cm}
\times
C_{12}(q_1^2,p^2,q_2^2,M_W^2,M_W^2,M_W^2)
\nonumber\\
&&
+ 2 M_W^2 \Big[
C_1(p^2,q_1^2,q_2^2,M_W^2,M_W^2,M_W^2)
+C_1(p^2,q_2^2,q_1^2,M_W^2,M_W^2,M_W^2)
\Big]
\nonumber\\
&&
+ 8 M_W^2 c_W^2 (s_W^2 - c_W^2) 
C_0(p^2,q_1^2,q_2^2,M_W^2,M_W^2,M_W^2)
\Bigg\}
, 
\end{eqnarray}
\begin{eqnarray}
F_{21}^{(G_3)}(p^2; q_1^2, q_2^2) 
&=& 
-\dfrac{e^3}{32 \pi^2 M_W s_W^3 c_W^4}
\times
\nonumber   \\
&&
\times
\Bigg\{
2 M_W^2
\Big[
C_1(p^2,q_1^2,q_2^2,M_Z^2,M_Z^2,M_H^2)
+C_1(p^2,q_2^2,q_1^2,M_Z^2,M_Z^2,M_H^2)
\Big]
\nonumber\\
&&
+(c_W^2 M_H^2+2 M_W^2) 
C_{12}(q_1^2,p^2,q_2^2,M_H^2,M_Z^2,M_Z^2)
\nonumber\\
&&
+3 c_W^2 M_H^2 
C_{12}(q_1^2,p^2,q_2^2,M_Z^2,M_H^2,M_H^2)
\Bigg\}.
\end{eqnarray}
The last form factor $F_{22}$ can be 
derived directly as 
\begin{eqnarray}
F_{22}(p^2; q_1^2, q_2^2) 
&=& 
F_{11}(p^2; q_2^2, q_1^2).
\end{eqnarray}
Moreover, it is easy to check that
all form factors satisfy the Bose
symmetry relations
\begin{eqnarray}
 F_{00, 12, 21}(p^2; q_1^2, q_2^2) 
&=& 
F_{00, 12, 21}(p^2; q_2^2, q_1^2).
\end{eqnarray}

We turn our attention to 
one-loop amplitude for off-shell
$H^*\rightarrow ZZ$.
In this case, we 
only have $F_{00}, F_{21}$ 
contributing to the amplitude. 
Analytic expressions
for these form factors can be 
obtained by taking $q_1^2=q_2^2=M_Z^2$. 
One-loop off-shell decay rates for 
$H^* \rightarrow ZZ,\; Z_LZ_L$ are computed. 
We use 
the following
kinematic variables: $p^2 = M_{ZZ}^2, 
q_1^2 = M_Z^2$ and $q_2^2 = M_Z^2$. 
Decay rate for the case of unpolarized
$Z$ bosons in final state gets the form of
\begin{eqnarray}
\Gamma_{H^* \rightarrow ZZ}
&=&
\dfrac{g_{HZZ}^2\; 
\sqrt{\lambda
\Big(M_{ZZ}^2,M_Z^2,M_Z^2\Big)
}
}{(64\pi)  M_Z^4 M_{ZZ}^3}
\Bigg\{
\; 
\Big(12 M_Z^4-4 M_Z^2 M_{ZZ}^2+M_{ZZ}^4\Big) +
\nonumber\\
&&\hspace{0.0cm}
+
\Big(2 M_{ZZ}^4
-8 M_Z^2 M_{ZZ}^2
+24 M_Z^4 \Big)
\mathcal{R}e\Big[
F_{00}(M_{ZZ}^2,M_Z^2,M_Z^2)
\Big]
\nonumber\\
&&\hspace{0cm}
+ 
\Big(8 M_Z^4 M_{ZZ}^2 
-6 M_Z^2 M_{ZZ}^4 
+ M_{ZZ}^6 \Big)
\mathcal{R}e\Big[
F_{21}(M_{ZZ}^2,M_Z^2,M_Z^2)
\Big]
\Bigg\}. 
\end{eqnarray}
Here, the Kall\"en function is defined as 
$\lambda (x; y, z) = (x - y - z)^2 - 4 yz$.

We next consider the polarized
$Z$ bosons. In rest frame of Higgs boson,
the longitudinal
polarization vectors for $Z$ bosons
are defined as: 
\begin{eqnarray}
 \varepsilon_{\mu}(q_i, \lambda=0) 
 = \dfrac{4 M_{ZZ}^2\; q_{i, \mu} -2 M_Z^2\; p_{\mu} }
 {M_Z \sqrt{\lambda(4 M_{ZZ}^2,M_Z^2,M_Z^2)} }, 
 \quad \text{for} \quad i=1,2.
\end{eqnarray}
By deriving again the squared amplitude for 
off-shell decay $H^* \rightarrow
Z_LZ_L$, we then arrive at 
\begin{eqnarray}
\Gamma_{H^* \rightarrow Z_L Z_L}
&=&
\dfrac{g_{HZZ}^2\; \sqrt{\lambda
\Big(M_{ZZ}^2,M_Z^2,M_Z^2\Big)
}
}{(2\pi)  M_Z^4 M_{ZZ}^3 
\lambda^2\Big(4 M_{ZZ}^2,M_Z^2,M_Z^2\Big)}
\Bigg\{
\; 
2 M_{ZZ}^4
\Big(M_Z^4-6 M_Z^2 M_{ZZ}^2+2 M_{ZZ}^4\Big)^2 +
\nonumber\\
&&
+ 
M_{ZZ}^4 
\Big(M_Z^4-6 M_Z^2 M_{ZZ}^2+2 M_{ZZ}^4\Big)
\times
\\
&&\hspace{2.9cm} \times
\Bigg[
\Big(4 M_Z^4-24 M_Z^2 M_{ZZ}^2+8 M_{ZZ}^4\Big)
\mathcal{R}e\Big[
F_{00}(M_{ZZ}^2,M_Z^2,M_Z^2)
\Big]
\nonumber\\
&&\hspace{0cm}
+ \Big(25 M_Z^4 M_{ZZ}^2
-20 M_Z^2 M_{ZZ}^4+4 M_{ZZ}^6\Big)
\mathcal{R}e\Big[
F_{21}(M_{ZZ}^2,M_Z^2,M_Z^2)
\Big
]
\Bigg]
\Bigg\}.
\nonumber
\end{eqnarray}
\section{Phenomenological results}%
\begin{figure}[ht]
\centering
$\begin{array}{cc}
\hspace{-5.2cm}
\Gamma_{H^*\rightarrow ZZ} [\text{GeV}] &
\hspace{-5cm}
\Gamma_{H^*\rightarrow ZZ} [\text{GeV}]
\\
\includegraphics[width=7cm,height=4.5cm]
{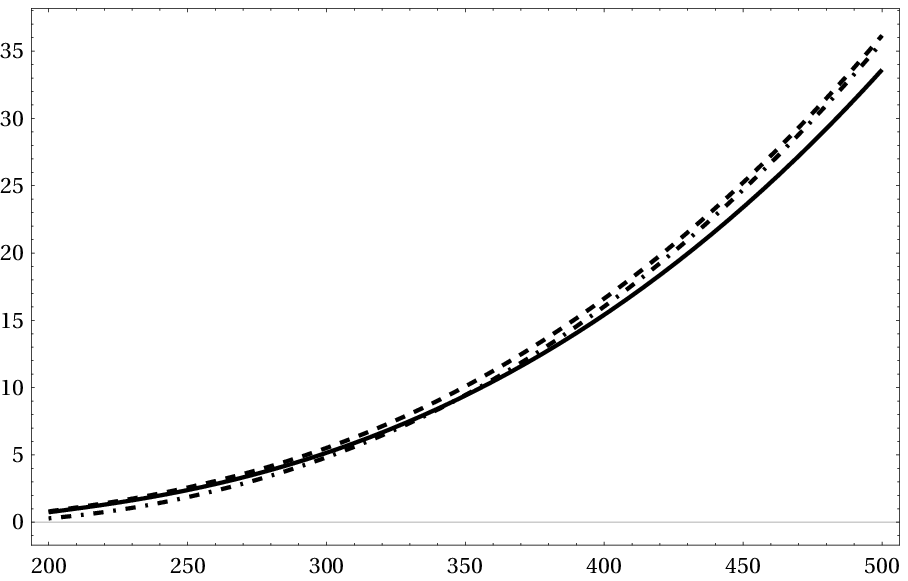}
& 
\includegraphics[width=7cm,height=4.5cm]
{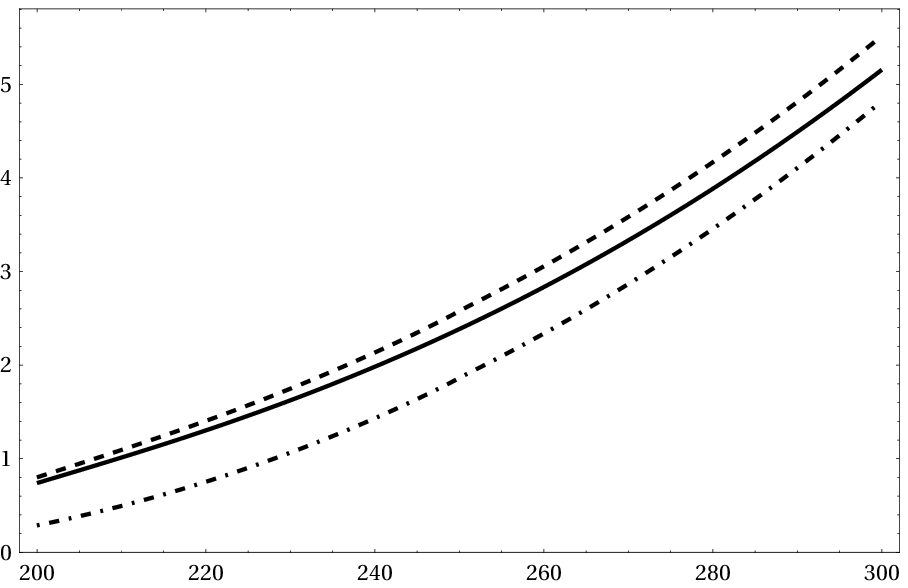}
\\
\hspace{5.2cm} M_{ZZ} [\text{GeV}]
&
\hspace{5.2cm} M_{ZZ} [\text{GeV}]
\end{array}$
\caption{\label{decay500} Off-shell Higgs 
decay rates as a function of 
$M_{ZZ}$.}
\end{figure}

In the phenomenological results, we use 
$M_Z = 91.1876$ GeV, 
$\Gamma_Z  = 2.4952$ GeV, 
$M_W = 80.379$ GeV, $\Gamma_W  = 2.085$ GeV, 
$M_H =125$ GeV, $\Gamma_H =4.07\cdot 10^{-3}$ GeV. 
The lepton masses are given: $m_e =0.00052$ GeV,
$m_{\mu}=0.10566$ GeV and $m_{\tau} = 1.77686$ GeV.
For quark masses, one takes $m_u= 0.00216$ GeV
$m_d= 0.0048$ GeV, $m_c=1.27$ GeV, $m_s = 0.93$ GeV, 
$m_t= 173.0$ GeV, and $m_b= 4.18$ GeV. 
We work in the so-called $G_{\mu}$-scheme
in which the Fermi constant 
is taken $G_{\mu}=1.16638\cdot 10^{-5}$ 
GeV$^{-2}$ and the 
electroweak coupling can be calculated 
appropriately
as follows:
\begin{eqnarray}
 \alpha = \sqrt{2}/\pi G_{\mu} M_W^2(1-M_W^2/M_Z^2)
 =1/132.184.
\end{eqnarray}
We then present the phenomenological results
in the following subsections. 
In Fig.~\ref{decay500}, off-shell 
Higgs decay rates as a function of $M_{ZZ}$
are shown. We vary $M_{ZZ}$ from $200$ GeV 
to $500$ GeV. 
In the left (right) panel of Fig.~\ref{decay500}, 
the decay rates are generated 
in the region of 
$200 \leq M_{ZZ}\leq 500$ GeV 
(and zoom out in $200 \leq M_{ZZ}\leq 300$ GeV 
to study the effects of $H^*\rightarrow Z_L Z_L$),
respectively. 
In these figures, the solid line presents 
for tree-level decay rates,
the dashed line shows for full one-loop 
electroweak decay rates. 
While the dash-dotted line is for 
full one-loop decay rates
with the longitudinal polarization 
for $Z$ bosons. We find that 
the decay rates in 
$H^*\rightarrow Z_L Z_L$ give 
small contributions 
in the low region of $M_{ZZ}$ 
and they tend to one-loop decay 
rates in high region
of $M_{ZZ}$.  

\begin{figure}[H]
\centering
$\begin{array}{cc}
\hspace{-6.5cm}
\delta [\%] &
\hspace{-6.5cm}
\delta [\%]
\\
\includegraphics[width=7cm,height=4.5cm]
{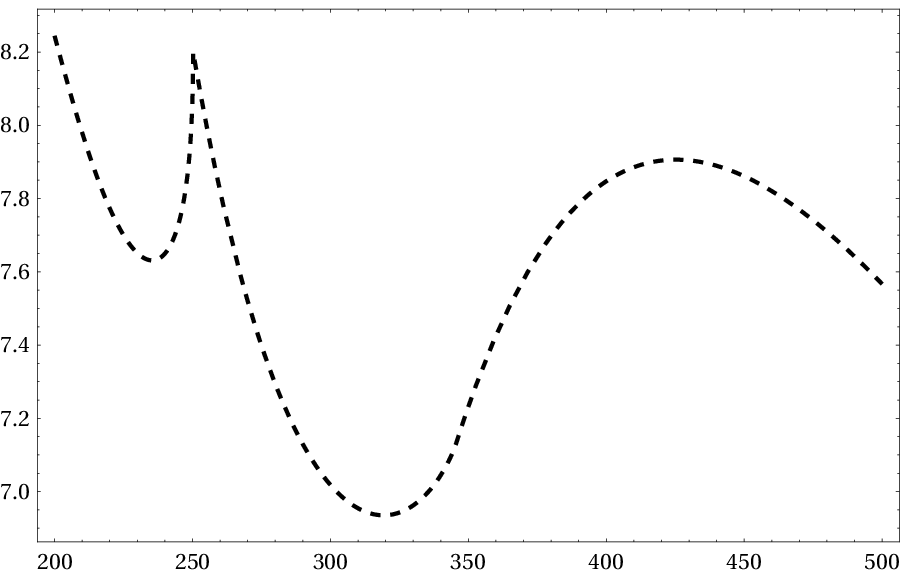}
& 
\includegraphics[width=7cm,height=4.65cm]
{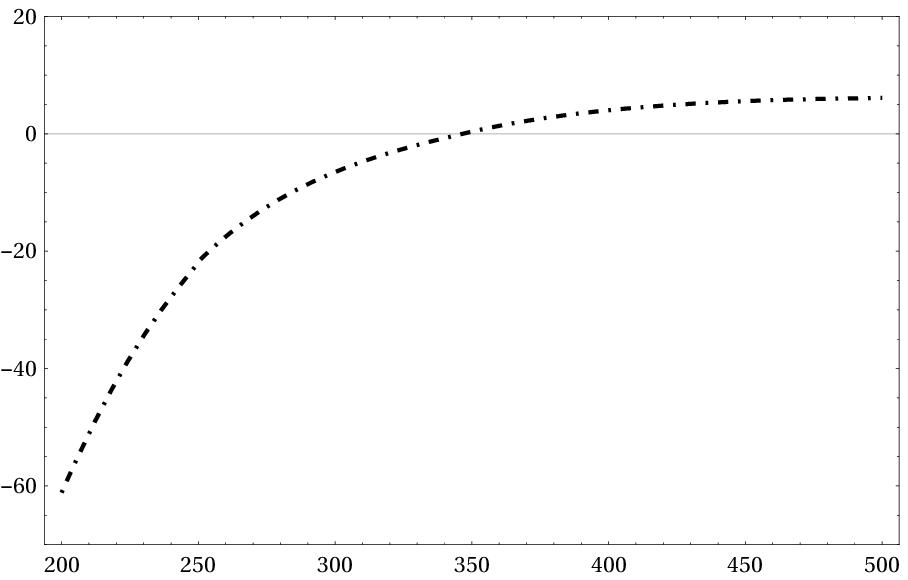}
\\
\hspace{5.2cm} M_{ZZ} [\text{GeV}]
&
\hspace{5.2cm} M_{ZZ} [\text{GeV}]
\end{array}$
\caption{\label{correctionDR} One-loop 
electroweak corrections to the 
decay rates as a function of $M_{ZZ}$.}
\end{figure}

In Fig.~\ref{correctionDR}, we show 
one-loop electroweak corrections to the 
decay rates. The corrections are defined
as follows:
\begin{eqnarray}
 \delta[\%] =\dfrac{\Gamma_{H^*\rightarrow 
 ZZ}^{\textrm{one-loop}} 
 -\Gamma_{H^*\rightarrow ZZ}^{\textrm{tree}}
 }{\Gamma_{H^*\rightarrow ZZ}^{\textrm{tree}}}
 \times 100\%. 
\end{eqnarray}
In the left panel, one presents one-loop
corrections for the case of unpolarized bosons
in final state. While the right figure shows 
for one-loop electroweak
corrections for the case of longitudinal 
polarization for both $Z$ bosons. 
We find that the corrections are in range 
of $7\%$ to $8.4\%$ for the case of 
unpolarized bosons. 
While the corrections change 
from $-60\%$ to $+10\%$ in the case of 
longitudinal polarization
for both $Z$ bosons. The effects
of one-loop electroweak corrections to 
off-shell Higgs decay $H^* \rightarrow ZZ$
are significant and
they should be taken into account
at future colliders. 

\begin{figure}[H]
\centering
$\begin{array}{cc}
\hspace{-6.5cm}
\sigma [\text{fb}] &
\hspace{-6.5cm}
\sigma [\text{fb}]
\\
\includegraphics[width=7cm,height=4.5cm]
{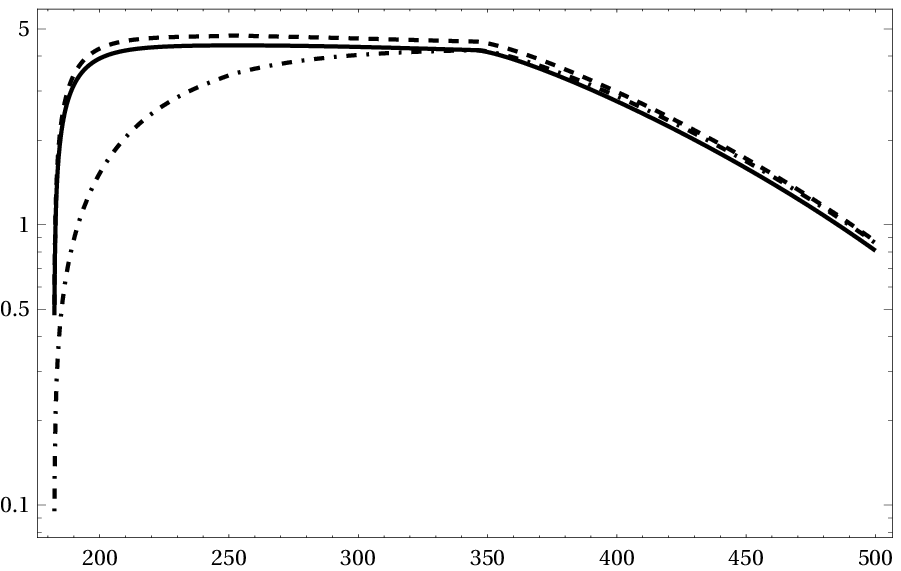}
& 
\includegraphics[width=7cm,height=4.65cm]
{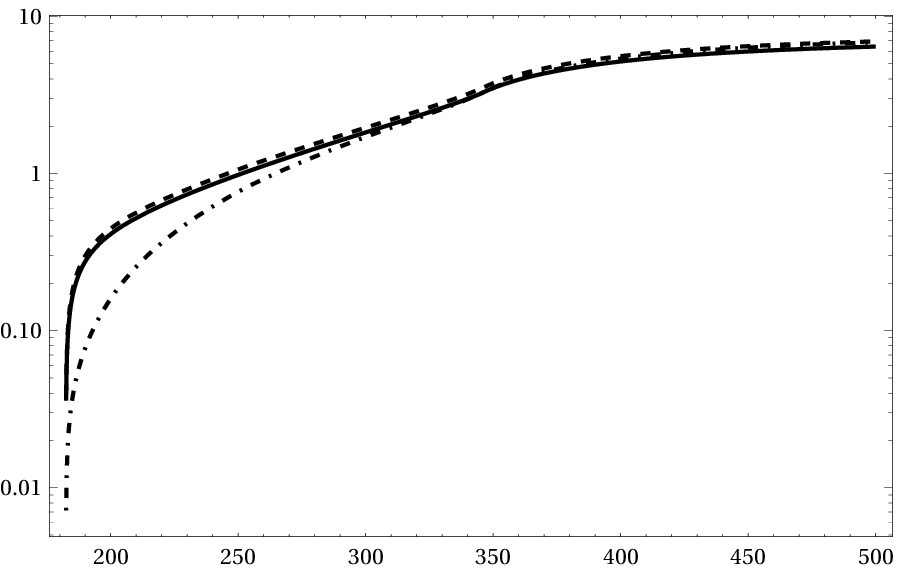}
\\
\hspace{5.2cm} \text{C.o.M} [\text{GeV}]
&
\hspace{5.2cm}\text{C.o.M} [\text{GeV}]
\end{array}$
\caption{\label{AA_H_ZZ} Total cross sections
are plotted as a function of center-of-mass
energy (C.o.M) for $Q^2=0$ (left panel) and 
$Q^2=1.5 M_H^2$ (right panel) respectively.}
\end{figure}

We study the effects of one-loop off-shell 
$H^* \rightarrow ZZ$ in Higgs 
productions at future colliders. The first 
signal process is $\gamma(Q^2)\gamma 
\rightarrow H^* \rightarrow ZZ$. The signal
cross section is given by 
\begin{eqnarray}
\sigma (\sqrt{s},Q^2)
&=&
\dfrac{2\sqrt{s}\;\; \Gamma_{H^* \rightarrow ZZ}
}{[(s-M_H^2)^2+ \Gamma_H^2
M_H^2] \sqrt{\lambda\big(s, Q^2, 0\big)} }
\Big|F_{00}^{H^*\rightarrow\gamma^*\gamma}
\big(s, Q^2, 0 \big)\Big|^2.
\end{eqnarray}
Where $F_{00}^{H^*\rightarrow\gamma^*\gamma}
\big(s, Q^2, 0 \big)$ is one-loop
form factor for process $H^* \rightarrow 
\gamma^*(Q^2)\gamma$ which its 
analytical result can be 
found in~\cite{Phan:2021ywd}. 
In Fig.~\ref{AA_H_ZZ}, total cross sections
are plotted as a function of center-of-mass
energy (C.o.M) for $Q^2=0$ (left panel) and 
$Q^2=1.5 M_H^2$ (right panel) respectively. 
In these Figures, the solid line shows for
tree-level cross sections, the dashed line 
is for one-loop contributing to 
$H^* \rightarrow ZZ$
and the dash-dotted line presents for 
one-loop contributing to 
$H^* \rightarrow Z_LZ_L$. 

\begin{figure}[H]
\centering
$\begin{array}{cc}
\hspace{-3.5cm}
d\sigma /d M_{ZZ} 
[\text{fb}/\text{GeV}]
&
\hspace{-3.5cm}
d\sigma/d Q^2 [\text{fb}/\text{GeV}^2]
\\
\includegraphics[width=7cm,height=4.5cm]
{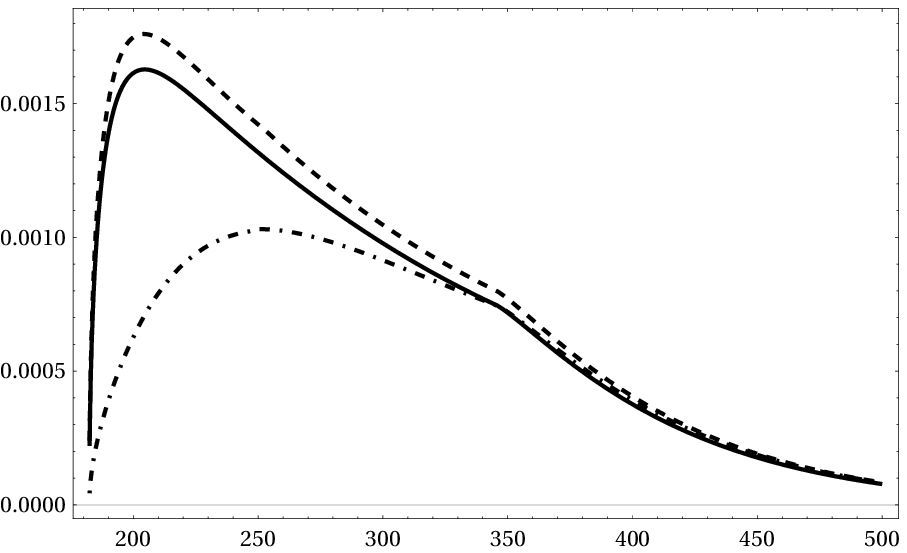}
& 
\includegraphics[width=7cm,height=4.65cm]
{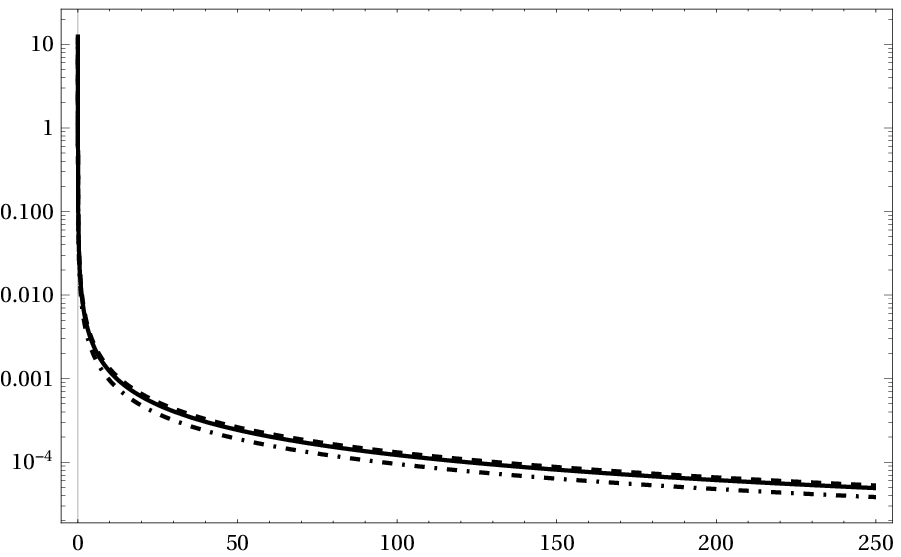}
\\
\hspace{5.2cm} M_{ZZ} [\text{GeV}]
&
\hspace{5.2cm} Q^2 [\text{GeV}]
\end{array}$
\caption{\label{AE_H_EZZ} .}
\end{figure}

The second signal process mentioned in this work 
is $e^-\gamma 
\rightarrow e^-H^* \rightarrow  e^-ZZ$. The signal
cross section is written as follows
\begin{eqnarray}
\dfrac{d^2 \sigma (\sqrt{s},Q^2)}{d M_{ZZ} \, d Q^2}
&=&
\dfrac{e^2}{16\pi s}
\Bigg[
\dfrac{s^2+(M_{ZZ}^2 -Q^2-s)^2}
{ Q^2(s^2-Q^2)^2}
\Bigg]\times \Big|F_{00}^{H^*\rightarrow\gamma^*\gamma}
\big(s, Q^2, 0 \big)\Big|^2
\times 
\nonumber\\
&& \times 
\dfrac{2 M_{ZZ}}
{[(M_{ZZ}^2-M_H^2)^2+ \Gamma_H^2
M_H^2 ]}\times 
\dfrac{ M_{ZZ}\; 
\Gamma_{H^*\rightarrow ZZ}
(M_{ZZ})}{\pi}. 
\end{eqnarray}
In Fig.~\ref{AE_H_EZZ}, we present
differential cross sections with respect
to off-shell Higgs mass $M_{ZZ}$ 
(left panel) and $Q^2$ (right panel).
In these distributions, we use the same 
previous notations. In the left Figure,
cross section develops to the peak  
which is corresponding to $M_{ZZ} \sim 2M_Z$.  
It then decrease rapidly beyond the peak.
It is interesting to observe that one-loop 
off-shell Higgs decay impacts are visible
around the peak. In the right Figure, 
we find that cross section is dominant 
in the low $Q^2$ regions. 
In both cases, one finds that
the effects of 
one-loop contributions to  off-shell
Higgs decay to $Z$-pair are visible
and they should be taken into account
at future colliders. 
\section{Conclusions}   
In this paper, we have performed  
one-loop electroweak contributing to 
$HZZ$ vertex in 't Hooft-Veltman gauge. 
We have also presented one-loop 
formulas for off-shell decay 
$H^* \rightarrow ZZ, Z_LZ_L$. Analytic 
expressions for one-loop form factors 
are shown in terms of the PV-functions 
in the standard notations of 
{\tt LoopTools}. Therefore, 
off-shell decay rates can be computed 
numerically by using this package.
One-loop electroweak corrections 
to the off-shell decay
rates are investigated for the cases of
unpolarized $Z$ bosons and longitudinal
polarization of $Z$ bosons in final state.
The corrections are range of $7\%$ 
to $8.4\%$ when varying off-shell Higgs
mass $200$ GeV $\leq M_{ZZ}\leq 500$ GeV.  
In applications, we study 
off-shell Higgs decay $H^* \rightarrow ZZ$
in the Higgs productions at 
future colliders such as the signal 
processes $\gamma^*(Q^2)\gamma 
\rightarrow H^* \rightarrow ZZ$
and 
$e^-\gamma 
\rightarrow e^-H^* \rightarrow e^- ZZ$
are studied. We find that
the effects of 
one-loop contributions to  off-shell
Higgs decay to $Z$-pair are visible
and they should be taken into account
at future colliders. 
\\

\noindent
{\bf Acknowledgment:}~
This research is funded by Vietnam National
University, Ho Chi Minh City (VNU-HCM) 
under grant number C$2022$-$18$-$14$.\\ 
\section*{Appendix A: Checks for the 
calculation}    
Before representing the phenomenological results, 
we are going to check the $UV$-finiteness of the 
results. As we mentioned in the previous 
section, the form factors $F_{00}^{(G_j)}$
for $j=1,2,3$ contain the $UV$-divergent. 
By taking the counter-term form factor 
$F_{00}^{(G_0)}$, the total form factor
$F_{00}$  is then $UV$-finite. 
The numerical results for this check
are presented in the following
Table~\ref{UV}. 
By changing
$C_{UV}, \mu^2$, we verify that the total 
form factor $F_{00}$ is very 
good stability (over $11$ digits). 
\begin{table}[ht]
\begin{center}
\begin{tabular}{l@{\hspace{2cm}}l}  
\hline \hline 
$(C_{UV}, \mu^2)$
& $\sum\limits_{j=1}^{3}F_{00}^{(G_j)}$\\
& $F_{00}^{(G_0)}$\\
&$F_{00}=\sum\limits_{j=0}^{3}F_{00}^{(G_j)}$  
\\ \hline \hline\\
$(0, 1)$ & $ -13.315051817147474 + 2.5760460657959703\; i$ \\
& $16.425138633496918 + 0\;i$ \\
& $3.110086816349444 + 2.5760460657959703\; i  $\\ \hline \\
$(10^2, 10^4)$ & $137.93892691326275 + 2.5760460657959703\; i $ \\
& $-134.82884009691335+ 0\;i$\\
& $3.110086816349394 + 2.5760460657959703\; i$\\ \hline \\
 $(10^4, 10^8)$ & $13861.982684608432 
 + 2.5760460657959703\;i $ \\
 & $-13858.872597792075+ 0\;i$ \\ 
& $3.1100868163575797 + 2.5760460657959703\; i$ \\ 
\hline\hline
\end{tabular}
\caption{\label{UV} Checking for 
the UV-finiteness of the 
results at $M_{ZZ} = 250$ GeV ($p^2=M_{ZZ}^2$). 
In this case, two real 
bosons are considered in final state.}
\end{center}
\end{table}
\section*{Appendix B:
Decay width of off-shell 
$H^* \rightarrow ZZ^*
\rightarrow Z l\bar{l}$ and 
$H^* \rightarrow Z^*Z^*
\rightarrow l_1\bar{l}_1 l_2\bar{l}_2$ with $l_{1,2}=e, \mu, 
\nu_{e}, \nu_{\mu}, \nu_{\tau}$}    
We also include the leptons decay from $Z$ boson. 
Since we are interested in the off-shell Higgs decay 
to $ZZ$. It means that $p_H^2 \geq 4 M_Z^2$.
Consequently, one can apply 
resonant approximation. The decay rates for
$H\rightarrow ZZ^* \rightarrow Z l\bar{l}$ 
can be then
presented in a compact form as: 
\begin{eqnarray}
\Gamma_{H\rightarrow Z^* Z\rightarrow Z l\bar{l}}
&=& 
\int\limits_{4m_l^2}^{(M_{ZZ}-M_{Z})^2}
\dfrac{d q_1^2}{\pi}\;
\dfrac{M_Z\; \Gamma^l_{Z}}{(q_1^2 - M_{Z}^2)^2
+ M_{Z}^2 \Gamma_{Z}^2 }
\dfrac{g_{HZZ}^2 \sqrt{\lambda\Big(M_{ZZ}^2,q_1^2,M_Z^2\Big)} }{(64\pi)  M_Z^3 M_{ZZ}^3 q_1^2}
\times
\nonumber\\
&&\hspace{0cm}
\times
\Bigg\{
\Big[M_Z^4-2 M_Z^2 (M_{ZZ}^2-5 q_1^2)+(M_{ZZ}^2-q_1^2)^2\Big]
\\
&& \hspace{0.5cm}
+
\Big[2 M_Z^4-4 M_Z^2 (M_{ZZ}^2-5 q_1^2)
+2 (M_{ZZ}^2-q_1^2)^2\Big]
\mathcal{R}e 
\Big[
F_{00}(M_{ZZ}^2,q_1^2,M_Z^2)
\Big]
\nonumber\\
&&\hspace{0.5cm}
- \Big(M_Z^2-M_{ZZ}^2+q_1^2\Big) 
\times 
\nonumber\\
&&
\hspace{1cm} \times 
\Big[M_Z^4-2 M_Z^2 (M_{ZZ}^2+q_1^2)
+(M_{ZZ}^2-q_1^2)^2\Big]
\mathcal{R}e 
\Big[
F_{21}(M_{ZZ}^2,q_1^2,M_Z^2)
\Big]
\Bigg\}.
\nonumber
\end{eqnarray} 
Where $M_Z\Gamma^l_Z =
\dfrac{M_Z\; g^2 s_W^2 \big(a_l^2 + v_l^2\big)}{12\pi}$
is partial decay rate of $Z$ to lepton pair with 
$a_l = T_3^{f}/(2 s_W c_W)$ and 
$v_l = (T_3^{f}-2Q_f s_W^2)/(2 s_W c_W) $. 
Following zero width approximation (ZWA)
for $Z$ decay into leptons, we employ
\begin{eqnarray}
 \dfrac{1}{(q_1^2 - M_{Z}^2)^2
+ M_{Z}^2 \Gamma_{Z}^2 }
\rightarrow \dfrac{\pi}
{M_{Z}\Gamma_{Z}} \;  \delta(q_1^2 -M_{Z}^2 ). 
\end{eqnarray}
One then has 
\begin{eqnarray}
 \int\limits_{4m_l^2}^{(M_{ZZ}-M_{Z})^2}
d q_1^2\;
\delta(q_1^2 -M_{Z}^2 ) =1. 
\end{eqnarray}
As a result, we arrive at 
\begin{eqnarray}
 \Gamma_{H\rightarrow Z^* Z\rightarrow Z l\bar{l}}
 = \Gamma_{H\rightarrow Z Z}
 \times \text{BR}_{Z\rightarrow l\bar{l}}. 
\end{eqnarray}

We next consider leptons decay from 
both $Z$ bosons. Applying 
the resonant approximation, 
one-loop off-shell decay rates 
$H^* \rightarrow Z^*Z^*
\rightarrow l_1\bar{l}_1 l_2\bar{l}_2 $ 
read: 
\begin{eqnarray}
\Gamma_{H\rightarrow Z^* Z^*\rightarrow 4
\; \text{leptons}}
&=& \int\limits_{4m_{l_1}^2}^{M_{ZZ}^2}
\dfrac{d q_1^2}{\pi}
\dfrac{M_Z\; \Gamma^{l_1}_{Z}}
{(q_1^2 - M_{Z}^2)^2 + M_{Z}^2 \Gamma_{Z}^2}
\int\limits_{4m_{l_2}^2}^{(M_{ZZ}-\sqrt{q_1^2})^2}
\dfrac{d q_2^2}{\pi}
\dfrac{M_Z\; \Gamma^{l_2}_{Z}}
{(q_2^2 - M_{Z}^2)^2 + M_{Z}^2 \Gamma_{Z}^2}
\\
&& \hspace{0cm} \times
\dfrac{g_{HZZ}^2 \sqrt{\lambda\Big(M_{ZZ}^2,q_1^2,q_2^2\Big)}}
{(64\pi) M_Z^2 M_{ZZ}^3 q_1^2 q_2^2}
\Bigg\{
\Big[M_{ZZ}^4-2 M_{ZZ}^2 (q_1^2+q_2^2)+q_1^4+10 q_1^2 q_2^2+q_2^4\Big]
\nonumber \\
&&\hspace{0cm}
+
\Big[2 M_{ZZ}^4-4 M_{ZZ}^2 
(q_1^2+q_2^2)+2 q_1^4+20 q_1^2 q_2^2+2q_2^4\Big]
\mathcal{R}e 
\Big[
F_{00}(M_{ZZ}^2,q_1^2,q_2^2) 
\Big]
\nonumber\\
&&\hspace{0cm}
+ \Big(M_{ZZ}^2-q_1^2-q_2^2\Big) \times 
\nonumber\\
&&\hspace{1.8cm} \times 
\Big[M_{ZZ}^4-2 M_{ZZ}^2 
(q_1^2+q_2^2)+(q_1^2-q_2^2)^2\Big]
\mathcal{R}e 
\Big[
F_{21}(M_{ZZ}^2,q_1^2,q_2^2)
\Big]
\Bigg\}.
 \nonumber
\end{eqnarray}
With the help of ZWA, we arrive at 
\begin{eqnarray}
 \Gamma_{H\rightarrow 
 Z^* Z^*\rightarrow Z l_1\bar{l}_1 l_2\bar{l}_2}
 = \Gamma_{H\rightarrow Z Z}\times 
 \text{Br}_{Z\rightarrow l_1\bar{l}_1 }
 \times \text{Br}_{Z\rightarrow l_2\bar{l}_2 }. 
\end{eqnarray}
\section*{Appendix $C$:              
Counter-term for the vertex $HZZ$}   
Counter-term for the $HZZ$ vertex 
has general form as follows~\cite{Aoki:1982ed}:
\begin{eqnarray}
\label{F00G0}
F_{00}^{(G_0)}(p^2; q_1^2, q_2^2)
=
\big(
\delta Y
+
\delta G_2
+
\delta G_3
+
\delta G_Z
+
2 \delta Z_{ZZ}^{1/2}
+
\delta Z_H^{1/2}
\big)
\left\langle
ZZH
\right\rangle
,
\end{eqnarray}
where $\left\langle ZZH \right\rangle$ 
will refer to the tree-level 
expression of the above vertex. 
All renormalization constants can be found 
in \cite{Aoki:1982ed,Tran:2022fdb}.
\begin{figure}[ht]
\centering
\includegraphics[width=6cm, height=3cm]
{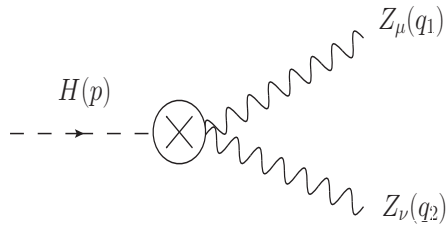}
\caption{Group $0$: counter-term Feynman
diagram.}
\end{figure}
\section*{Appendix $D$: Feynman diagrams}
\begin{figure}[ht]
\centering
\includegraphics[width=15cm, height=3cm]
{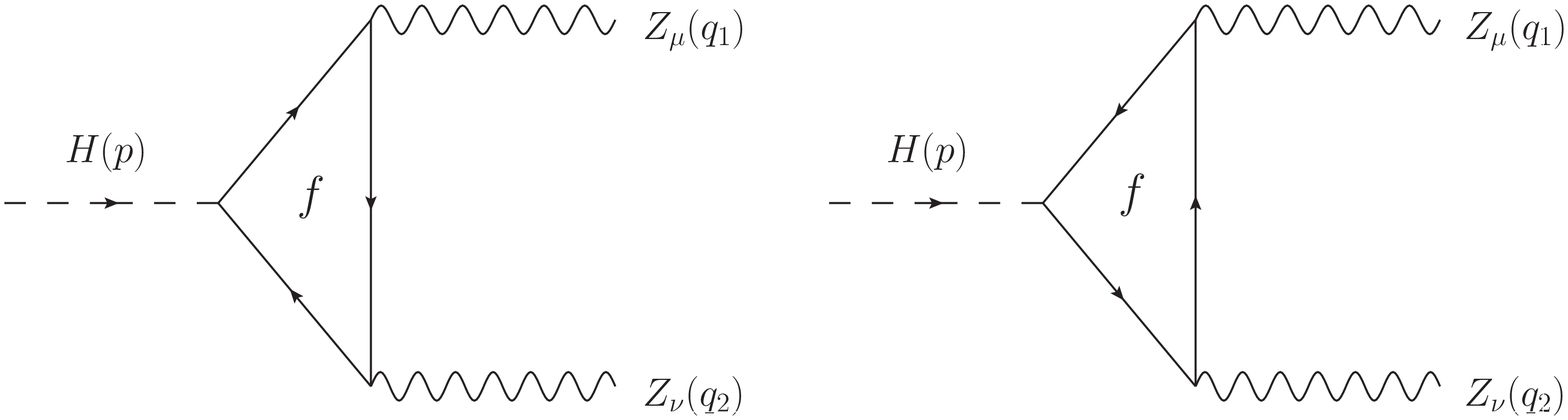}
\caption{one-loop Feynman
diagrams with exchanging $f$ 
in the loop (Group $1$).}
\end{figure}
\begin{figure}[ht]
\centering
\includegraphics[width=15.0cm, height=3.2cm]
{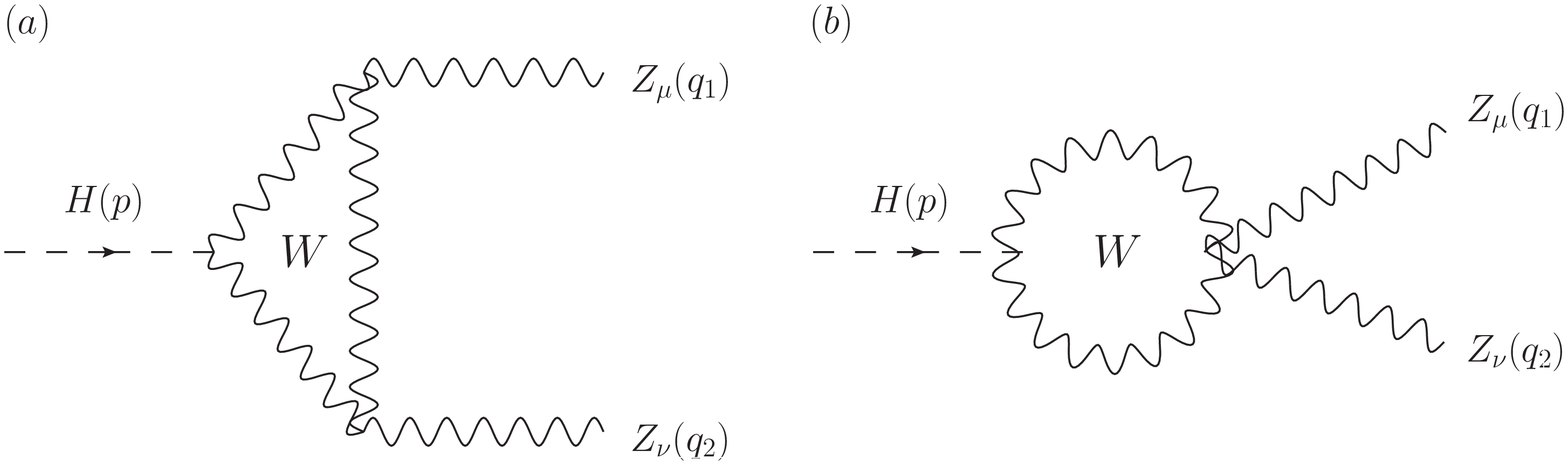}
\includegraphics[width=15.0cm, height=3.0cm]
{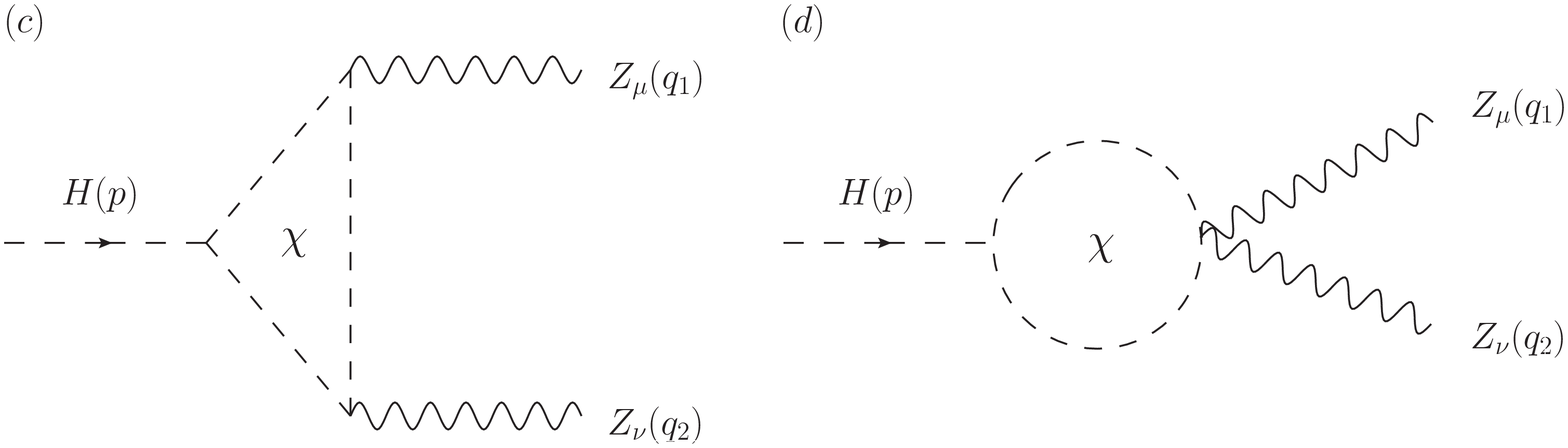}
\includegraphics[width=15.0cm, height=3.0cm]
{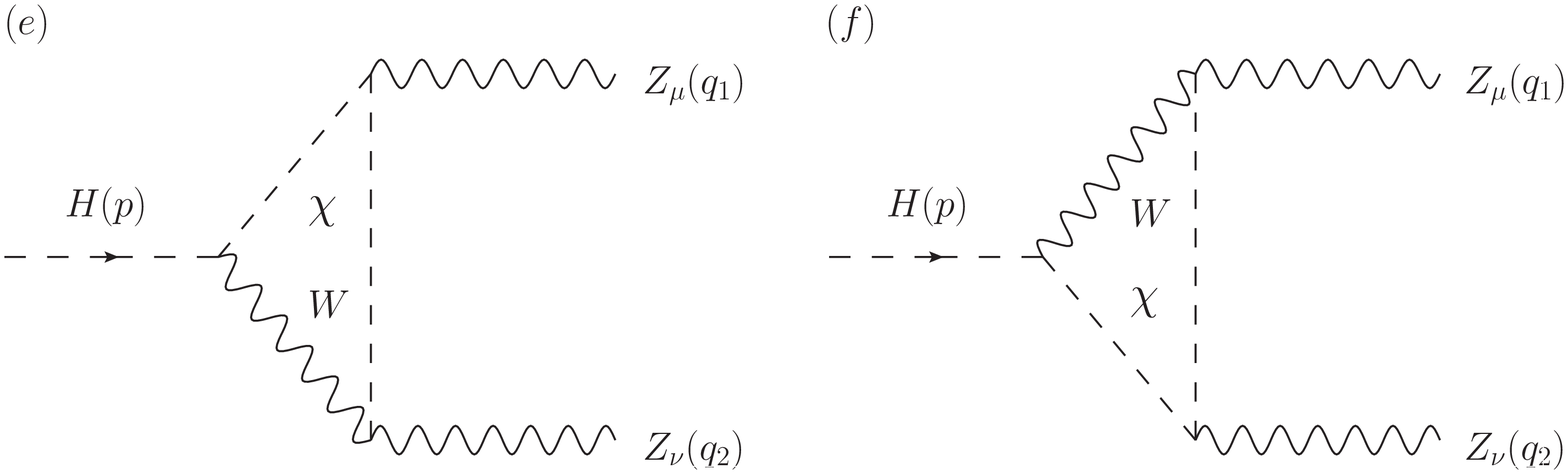}
\includegraphics[width=15.0cm, height=3.0cm]
{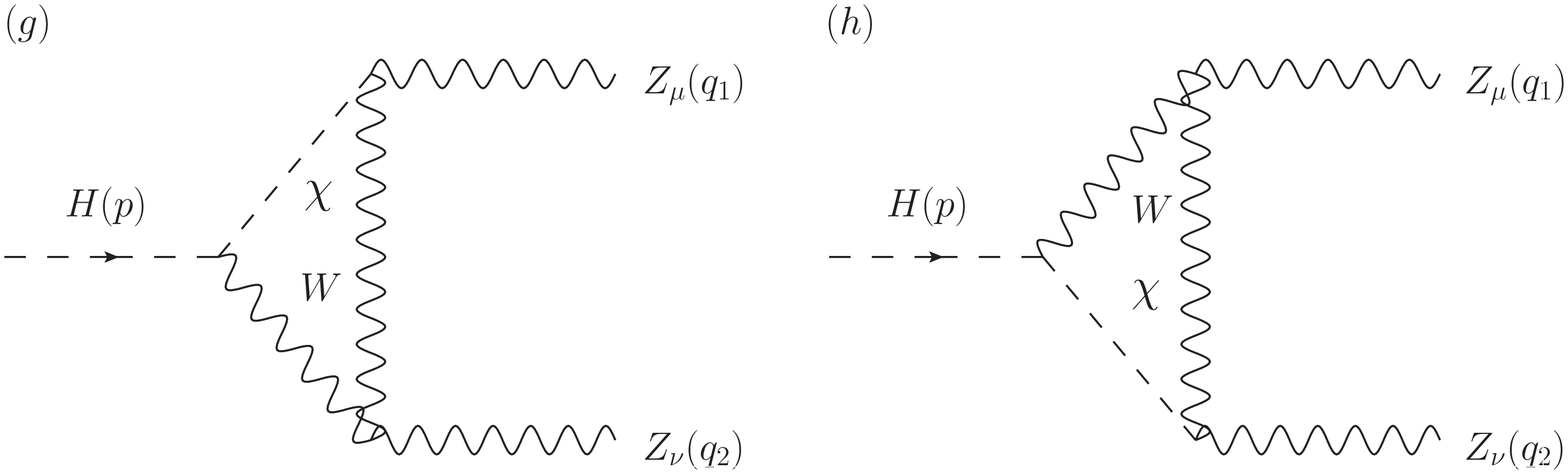}
\includegraphics[width=15.0cm, height=3.0cm]
{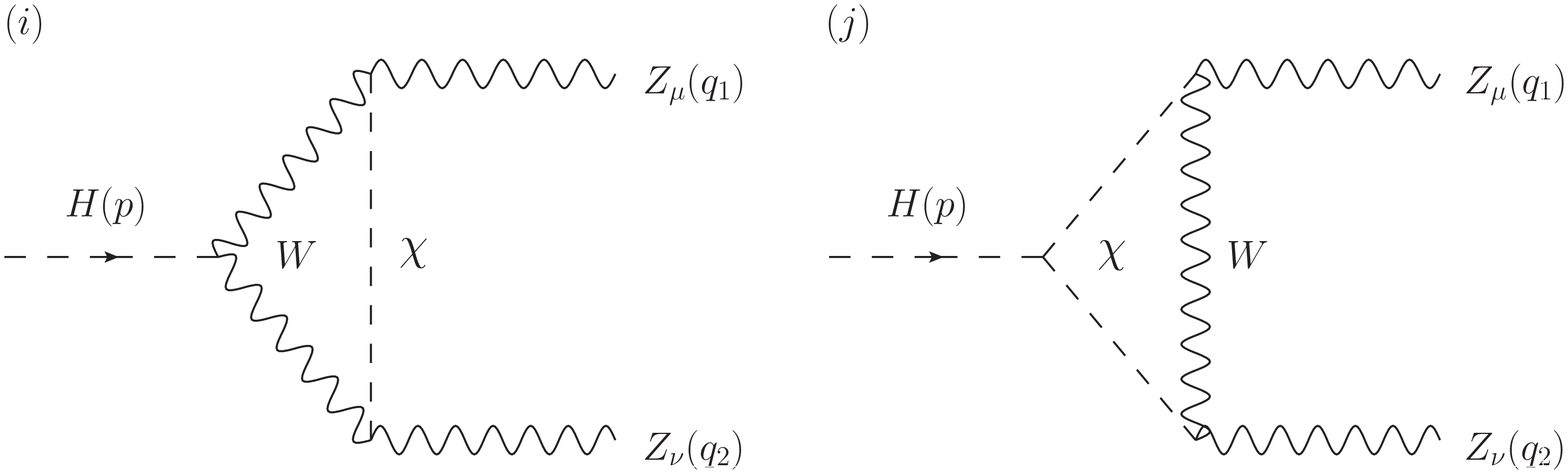}
\includegraphics[width=15.0cm, height=2.8cm]
{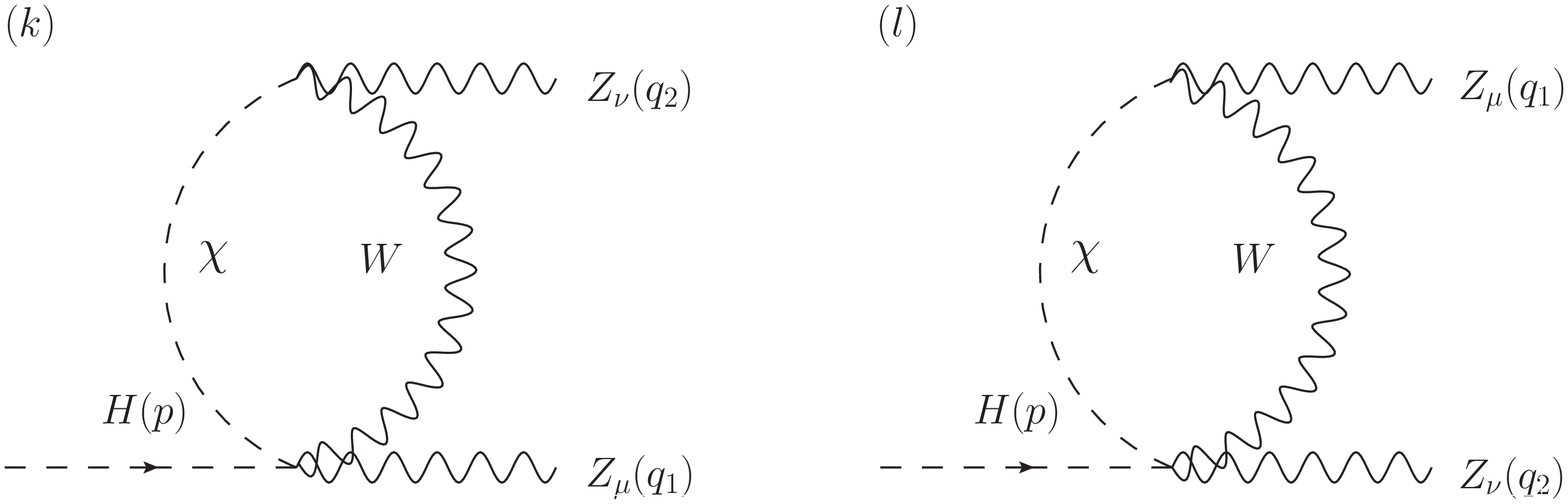}
\includegraphics[width=15.0cm, height=3.0cm]
{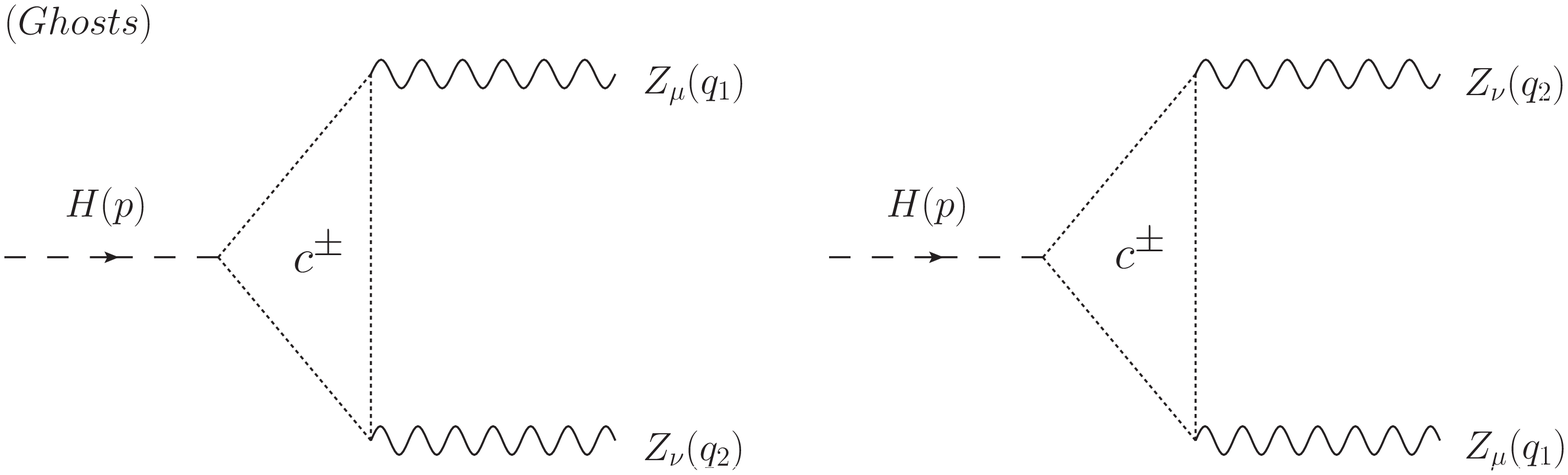}
\caption{one-loop Feynman
diagrams with exchanging $W, \chi$ 
and ghost particles
in the loop (Group $2$).}
\end{figure}
\begin{figure}[ht]
\centering
\includegraphics[width=15.0cm, height=3.0cm]
{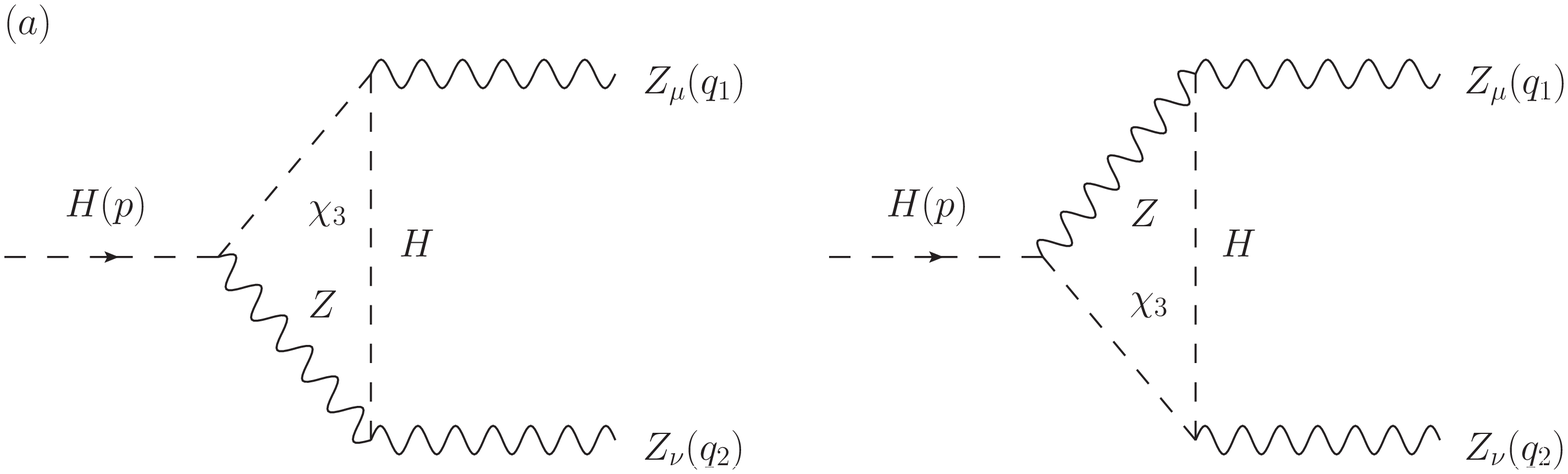}
\includegraphics[width=15.0cm, height=3.0cm]
{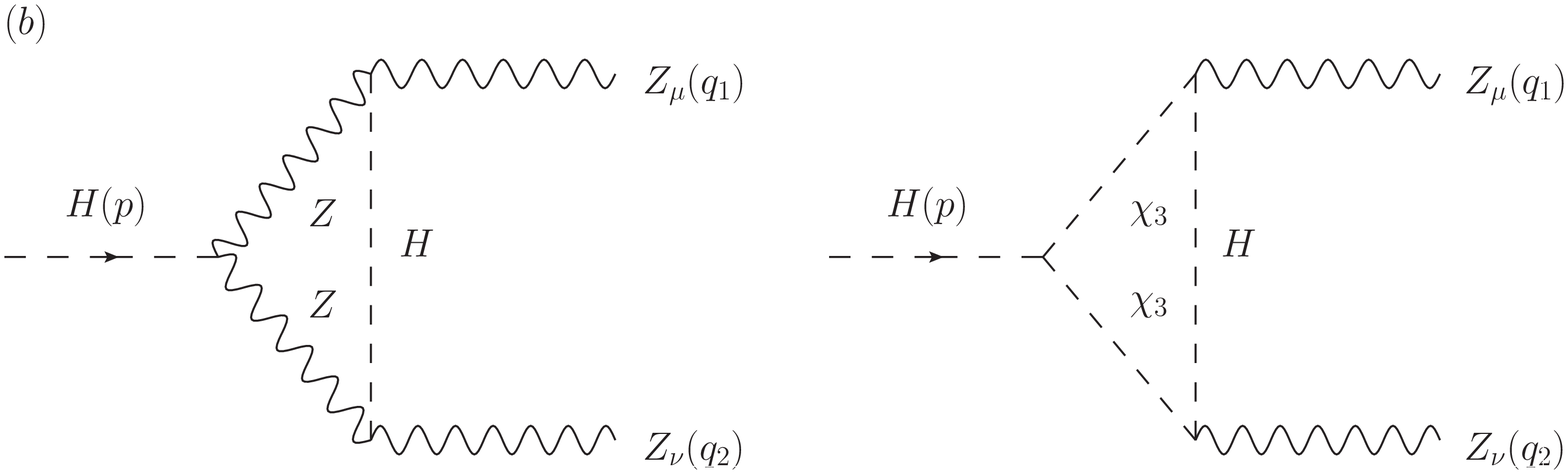}
\includegraphics[width=15.0cm, height=3.0cm]
{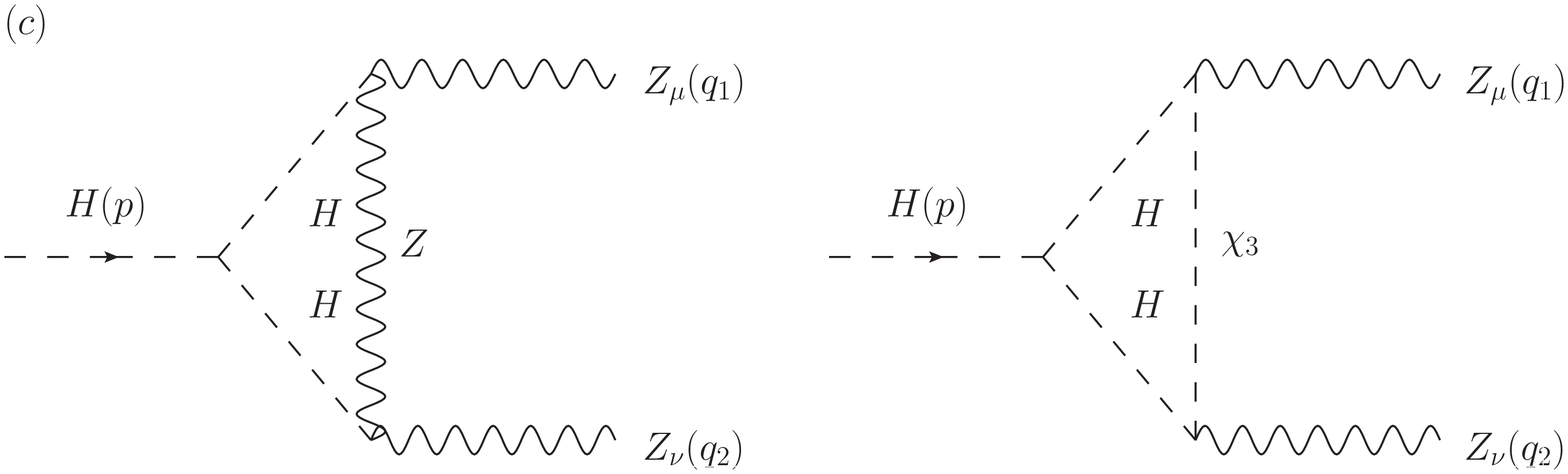}
\includegraphics[width=15.0cm, height=3.3cm]
{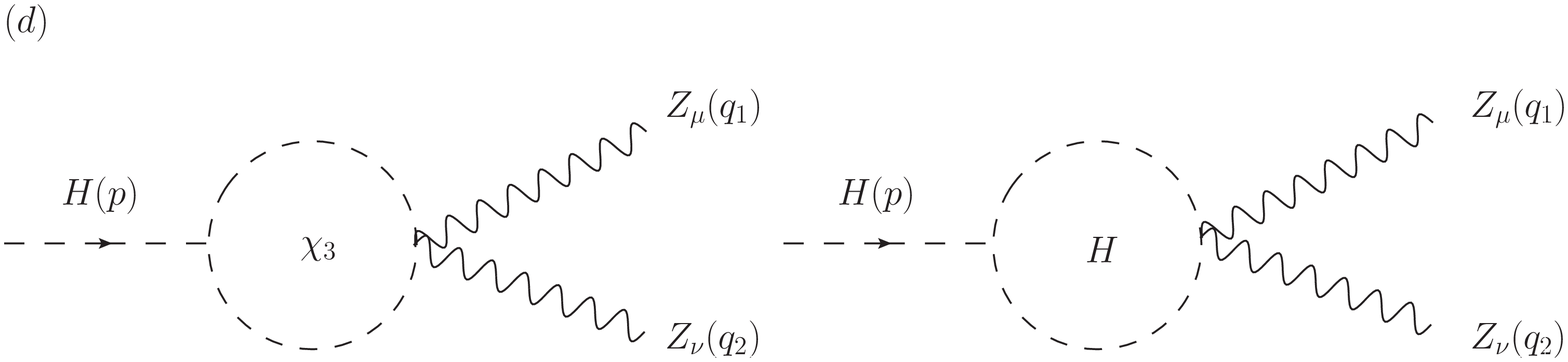}
\includegraphics[width=15.0cm, height=3.6cm]
{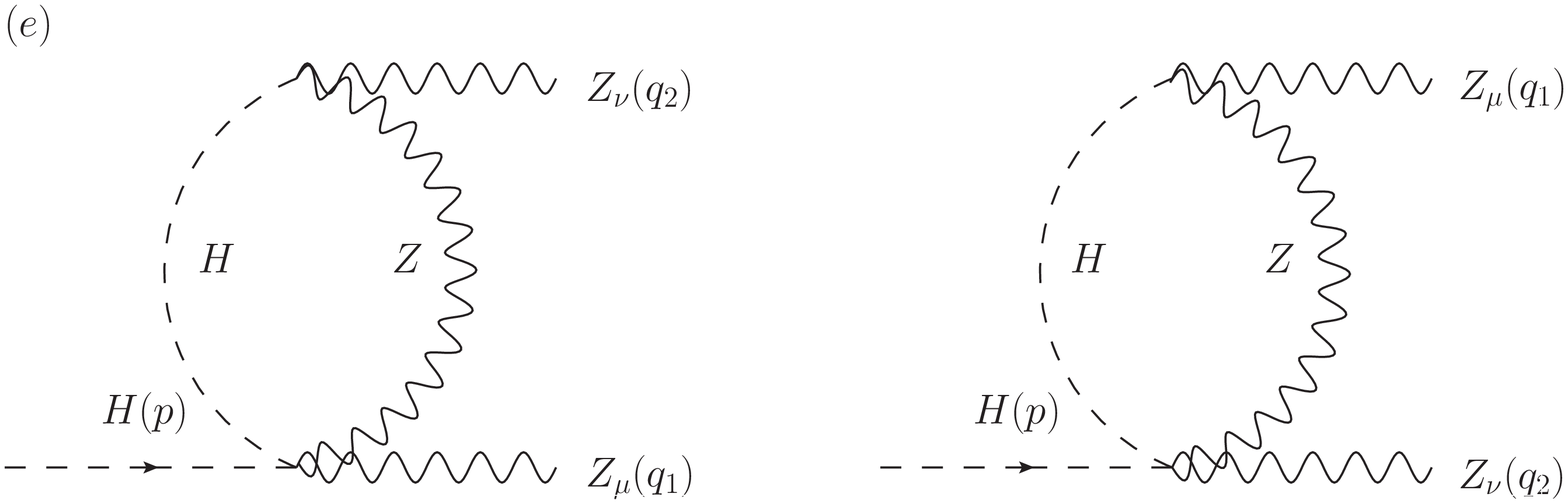}
\caption{one-loop Feynman
diagrams with exchanging $Z, \chi_3$ 
and $H$
in the loop (Group $3$).}
\end{figure}

\end{document}